\RequirePackage{fix-cm}
\documentclass[smallextended]{svjour3}       
\smartqed  
\usepackage[all]{xy}
\usepackage{graphicx}
\usepackage{algorithm, algorithmic, xcolor}%
\usepackage{cite}

\begin{document}

\title{Electrodiffusion with calcium-activated potassium channels in dendritic spine}


\author{Pilhwa Lee}


\institute{Pilhwa Lee \at
              Department of Mathematics, Morgan State University \\
              Baltimore, MD\\
              \email{pilhwa.lee@morgan.edu}           
           \and
}

\date{Received: date / Accepted: date}

\maketitle

\begin{abstract}

We investigate calcium signaling feedback through calcium-activated potassium channels of a dendritic spine by applying the immersed boundary method with electrodiffusion. We simulate the stochastic gating of such ion channels and the resulting spatial distribution of concentration, current, and membrane voltage within the dendritic spine. In this simulation, the permeability to ionic flow across the membrane is regulated by the amplitude of chemical potential barriers. With spatially localized ion channels, chemical potential barriers are locally and stochastically regulated. This regulation represents the ion channel gating with multiple subunits, the open and closed states governed by a continuous-time Markov process. The model simulation recapitulates an inhibitory action on voltage-sensitive calcium channels by the calcium-activated potassium channels in a stochastic manner as a \emph{non-local} feedback loop. The model predicts amplified calcium influx with more closely placed channel complexes, proposing a potential mechanism of differential calcium handling by channel distributions. This work provides a foundation for future computer simulation studies of dendritic spine motility and structural plasticity.

\end{abstract}

\section{Introduction}

\indent\indent Dendritic spines are small protrusions in the postsynapse and dendritic trees of neurons (Shepherd 1996). They are crucial in learning and memory (Harris and Kater 1993, Seung and Kim 2002, Yasuda et al. 2003, Matus 2004, Bloodgood and Sabatini 2007). In developmental stages, they are highly motile in the course of neuronal wiring and pruning (Bonhoeffer and Yuste 2002, Tada and Seung 2006, Bosch and Hayashi 2011). However, dendritic spine motility and potentially involving structural plasticity are persistently observed in matured stages (Lamprecht et al. 2006) and quite evident in aging and neurodegeneration, neuronal injury, as well as psychiatric disorders (Koleske 2013, Kim et al. 2015, Rice et al. 2015). In the presence of several ion channels in the dendritic spine (Sabatini and Svoboda 2000, Bloodgood and Sabatini 2007b), relevant modeling with electrodiffusion in the micro- and nano-domains is needful to reconstruct physiological ionic transports and understand consequential motility and associated neuronal functions (Holcman and Yuste 2015).

Previously we have studied the electrodiffusion with stochastic voltage-sensitive calcium channels in a one-dimensional setting using an immersed boundary method for electrodiffusion (Lee et al. 2013). The interface conditions for membrane permeability to each ionic species are replaced by chemical potential barriers in a unified Cartesian domain without explicit dissection of the computational domain between intracellular and extracellular domains. This article extends this approach to a two-dimensional domain considering the spatial effects of stochastic channel gating and channel distribution. We focus on voltage-sensitive calcium channels and calcium-activated potassium channels. In the configuration of the spatial expression of those two types of channels, they are mostly coupled together as ``complex" with about 15 nm distance in the nanodomain (Berkefeld et al. 2006). There are two types of calcium-activated potassium channels; small-conductance (SK) and large-conductance (BK) (Sah and Faber 2002). These potassium channels give feedback to the voltage-sensitive calcium channels with repolarization/hyperpolarization to inhibit further calcium influx (Griffith et al. 2016) and regulate in the synaptic strength and short-term plasticity among many other roles (He et al. 2014, Lujan et al. 2018, Gutzmann et al. 2019).

The chemical potential barriers are chosen so that the membrane is permeable to Ca$^{2+}$ and K$^+$ locally in space when the voltage-sensitive calcium channels and the calcium-activated potassium channels are open. To recapitulate the back-propagating action potential from the soma, the membrane is also regulated uniformly in space to be semi-permeable to Na$^+$ for the membrane depolarization. The membrane is uniformly semi-permeable to Cl$^-$ all the time of the simulation.

We place five voltage-sensitive calcium channels and five calcium-activated potassium channels (BK) in the dendritic spine head. According to continuous-time Markov processes, the opening and closing of the channels are modeled by lowering and raising the heights of the chemical potential barriers of calcium and potassium. The voltage-sensitive calcium channels have four independent subunits with inactivation from the intracellular local calcium concentration (Cox 2014). The channels are open only when all four subunits are in the open state. Voltage-sensitive rate constants govern the transitions between the open and closed states of the subunits, and the transition to the inactivated state is governed by the intracellular calcium concentration. The calcium-activated potassium channels also have four independent subunits with closed and open states dependent on the membrane voltage, and each subunit is activated by the intracellular calcium (Cox 2014).

As a whole, we have incorporated \emph{an immersed boundary method with electrodiffusion} for ionic transport in the dendritic spine of synapse on the two-dimensional domain, coupled to continuous-time Markov processes for the ion channel gating of voltage-sensitive calcium channels and calcium-activated potassium channels, possibly distributed non-uniformly in the dendritic spine. The model simulation recapitulates an inhibitory action on voltage-sensitive calcium channels by calcium-activated potassium channels in a stochastic manner as a \emph{non-local} feedback loop. The model also predicts enhanced calcium influx with more closely placed channel complexes, proposing a mechanism of differential calcium handling by the \emph{locality} of channel distribution.

The paper is organized in the following way; in Section 2, we describe the mathematical formulation of the electrodiffusion of ion species in the immersed boundary formalism, the ion-channel gating as a continuous-time Markov process, and the resulting regulation of the chemical potential barriers that model ion-channel selectivity. In Section 3, we present a two-dimensional study with spatially localized channels. In Section 4, we summary the model results and mention future works.

\section{MATHEMATICAL FORMULATION} \label{sec:mathematical_formulation}

\indent\indent In this section, we consider a fixed two-dimensional computational domain with dissolved ions. Immersed within the domain is a closed membrane, which is fixed in place. The membrane may be permeable or impermeable to each ionic species, the permeability being controlled in a graded manner by its chemical potential barrier. We have the following notations:\\

\noindent$D_{i}$: diffusion coefficient of the $i^{\rm th}$ ion species\\
$q$: the unit electrical charge (charge on a proton)\\
$qz_{i}$: charge of the $i^{\rm th}$ species\\
$K_{\rm B}$: Boltzmann constant\\
$T$: absolute temperature (degrees in Kelvin)\\
$\Omega_{\rm E}$: Eulerian domain\\
$\Omega_{\rm L}$: Lagrangian domain\\
$\varepsilon $: dielectric constant\\

\noindent The notations for the variables are the following:\\
$\mathbf{x}=(x_1,x_2)$: \noindent Eulerian coordinate\\
$\mathbf{x}=\mathbf{X}(s)$: Lagrangian description\\
$\psi_{i}(\mathbf{x}, t)$: chemical potential of the $i^{\rm th}$ ion species\\
$\Psi(\mathbf{x})$: chemical potential kernel\\
$A_{i}(s,t)ds$: contribution of arc $(s, s+ds)$ of the membrane to the chemical potential of $i^{\rm th}$ ion species\\
$c_{i}(\mathbf{x}, t)$: concentration of the $i^{\rm th}$ ion species\\
{$\mathbf{J}_i(\mathbf{x},t)$}: flux per unit area of the $i^{\rm th}$ ion species\\
$\varphi(\mathbf{x}, t)$: electrical potential\\
$\rho_0(\mathbf{x})$: background electrical charge density

\subsection{The chemical potential}

\indent\indent The chemical potential is expressed in the following way:
\begin{equation}
\psi_{i}(\mathbf{x},t) = \int_{\Omega_{\rm L}} \Psi(\mathbf{x} - \mathbf{X}(s), \mathbf{t}, \mathbf{n})A_i(s,t)ds.
\end{equation}
Here, $\mathbf{X}(s)$ is the configuration of the immersed boundary, where $s$ is a Lagrangian parameter. The function $A_i(s,t)$
describes the contribution of the membrane at $\mathbf{X}(s)$ to the chemical potential
barrier for the $i^{\rm th}$ ionic species. The chemical potential kernel $\Psi$ defines how the
contribution $A_i(s,t)ds$ is to be spread out in space in the neighborhood
of $\mathbf{X}(s)$. In regulating membrane permeability to the $i^{\rm{th}}$ species of ion, the chemical potential amplitude $A_i(s,t)$ is modulated locally or globally. When it is locally controlled, the localized area represents the domain of ion channels for the $i^{\rm{th}}$ ionic species.

In general, any bell-shaped function with compact support can be available for the chemical potential kernel. For the one-dimensional kernel, a smoothed Dirac delta function $\phi$ of the second-order moment with compact support is used following the function constructed by Peskin \cite{peskin}. The chemical potential kernel on the two-dimensional domain is as follows:
\begin{equation}
\Psi_{w}(\mathbf{x}, \mathbf{t}, \mathbf{n}) = \frac{1}{w^2} \phi(\frac{\mathbf{x} \cdot \mathbf{t}}{w})
\phi(\frac{\mathbf{x} \cdot \mathbf{n}}{w}),
\end{equation}
where $w$ is a scaling factor such that $\Psi_w$ has a support of a square of edge $4w$. The coordinates used in the $\phi$ are in the local frame from the tangential and normal directions, $\mathbf{t}$ and $\mathbf{n}$ concerning the membrane at $\mathbf{X}(s)$.

\subsection{The electrostatic potential and the electrodiffusion: Poisson-Nernst-Planck equations}

\indent\indent The electrical potential is a solution of the Poisson equation:
\begin{eqnarray}
-\nabla \cdot ( \epsilon \nabla \varphi ) = \sum_{i}qz_{i}c_{i} + \rho_0,  \label{eq:poisson1}
\end{eqnarray}
where $\rho_0$ represents the background electrical charge density. The two-dimensional domain is prescribed to be periodic in each direction, and the necessary condition for the existence of the solution of the Poisson equation requires the global electroneutrality. In the immersed boundary method with electrodiffusion, local electroneutrality is also satisfied except for the space charge layer around the membrane (Lee 2007, Lee et al. 2010). 

The electrodiffusion equations are formulated in the following way:
\begin{eqnarray}
\frac{\partial c_{i}}{\partial t} + \nabla \cdot \mathbf{J}_i & = & 0, \label{eq:continuity}\\
\mathbf{J}_{i} & = & -D_{i}(\nabla c_{i} + c_{i} \frac{\nabla
(\psi_{i} + qz_{i}\varphi)}{K_{\rm B}T}). \label{eq:flux}
\end{eqnarray}
Eq.(\ref{eq:continuity}) is the conservation law (the continuity equation) for the
$i^{\rm th}$ ionic species. In this equation, $c_i$ is the concentration, and $\mathbf{J}_{i}$ is
  the flux per unit area of this ion species.
Eq.(\ref{eq:flux}) gives the flux per unit area as a sum of three terms: diffusion, drift caused by the chemical potential, and drift caused by the electrical potential.
 
\subsection{Continuous-time Markov process for the stochastic ion channel gating} \label{section:continuous-time-Markov}

\indent \indent We will reformulate the kinetic models of voltage-sensitive calcium channels and calcium-activated potassium channels in Cox 2014 as continuous-time Markov processes to represent their stochastic channel gating activities on the spatial domain. For the ion channel with the center at $\mathbf{X}(s)$, the involving membrane voltage, $V_{\rm m}(s)$, the intracellular concentration, $c_i(s)$, and ionic current, $I_i(s)$ are collected from the following:
\begin{eqnarray}
V_{\rm m}(s) &=& \int_{\Omega_{\rm E}} (\Psi_w(\mathbf{x} -\mathbf{X}(s) + 2w \mathbf{n}) -  \Psi_h(\mathbf{x} -\mathbf{X}(s) - 2w \mathbf{n})) \varphi(\mathbf{x}) d\mathbf{x}, \\
c_i(s) &=& \int_{\Omega_{\rm E}} \Psi_w(\mathbf{x} - \mathbf{X}(s) + 2w \mathbf{n}) c_i(\mathbf{x}) d\mathbf{x}, \\
I_i(s) &=& qz_i \int_{\Omega_{\rm E}} \Psi_w(\mathbf{x} - \mathbf{X}(s) + 2w \mathbf{n}) \mathbf{J}_i(\mathbf{x}) \cdot \mathbf{n} d\mathbf{x},
\end{eqnarray}
where $4w$ is mostly the width of the membrane, and $\mathbf{n}$ is the normal unit vector at $\mathbf{X}(s)$ outward from the membrane.

\subsubsection{The voltage-sensitive calcium channel gating} 

The transition between closed and open states of each subunit is expressed as follows:
\begin{equation}
\xymatrix{{\rm C_S} \ar@<1ex>[r]^{\alpha(V_{\rm m})}& {\rm O_S} \ar@<1ex>[l]^{\beta(V_{\rm m})} }, \label{kinetics}
\end{equation}
where $\rm C_S$ and $\rm O_S$ represent closed and open states of a subunit. The rate constants of ${\alpha}$ and ${\beta}$ are functions of membrane voltage $V_{\rm m}$.\\

We describe the states of subunits and an ion channel in the on/off way:
\begin{eqnarray}
\chi_i &=& \left\{ \begin{array}{ll}
 1  &  {\rm subunit~ open} \\
 0	& {\rm subunit~ closed} \end{array} \right. \label{kinetics_s_i}\\
 S &=& \left\{ \begin{array}{ll}
 {\rm C}_{i-1}  & i-1 {\rm ~subunits~are~open} \\
 {\rm O} & {\rm ion~channel~open}\\
{\rm I} & {\rm inactivated}~{\rm state}
 \end{array} \right. \label{kinetics_p}
\end{eqnarray}
where $\chi_i$ indicates the open/closed state of $i^{\rm th}$ subunit, and $S$ the state of the ion channel. When we express the transition probability between the open and closed states of each subunit based on Eq.(\ref{kinetics}),
\begin{eqnarray}
&&P(\chi_i(t+dt)=1|\chi_i(t)=0) = \alpha(V_m)dt \label{closed_open}, \label{eq:transition1}\\
&&P(\chi_i(t+dt)=0|\chi_i(t)=1) = \beta(V_m)dt, \label{open_closed}\\
&&\alpha(V_m) = \alpha_0 e^{q_{\rm forward} V_{\rm m}/26.7},\\
&&\beta(V_m) = \beta_0 e^{-q_{\rm backward} V_{\rm m}/26.7},
\end{eqnarray}
where $\alpha_0$ = 3.0 ms$^{-1}$, $\beta_0$ = 0.241 ms$^{-1}$, $q_{\rm forward} = 1.16$,  and $q_{\rm backward} = 1.94$, where $V_{\rm m}$ is in the unit of mV. Eq.(\ref{closed_open}) represents the probability of $i^{\rm th}$ subunit to take the transition from the closed state at $t$ to the open state at $t+dt$ in the infinitesimal time interval $dt$.
For the individual ion channel gating, a continuous-time Markov process is applied \cite{peskin2}. The ion channel is assumed to have four independent subunits; each has open and closed states.

The channel is open only when all four subunits are in the open state, and when the channel is not inactivated. The diagram for the Markov process with the discrete states is presented in Fig.(\ref{markov_CaV}). In the discrete states labeled $C_i$,  $i-1$ represents the number of subunits in the open state. The state with all four subunits open, however, is given the special symbol O. The ion channel is open when all those subunits are open, i.e., when it is in the state O. In the state of I, the channel is inactivated and closed.

\subsubsection{The calcium-activated potassium channel gating}

The channel is also constituted with four independent subunits (Fig. \ref{markov_BK}). Each subunit is activated by calcium, with binding and unbinding rate constants $K_{\rm o} [{\rm Ca}^{2+}]$ and $K_{\rm -o}$ in the open state, and $K_{\rm c} [{\rm Ca}^{2+}]$ and $K_{\rm -c}$ in the closed state. The constants $K_{\rm o}$, $K_{\rm -o}$, $K_{\rm c}$, and $K_{\rm -c}$ are 1.0 nM$^{-1}$s$^{-1}$, 1.065 ms$^{-1}$, 1.0 nM$^{-1}$s$^{-1}$, 11.917 ms$^{-1}$, respectively. With $i$ subunits activated by calcium, the open and closed states are determined by the forward and backward rate constants, $K_i$ and $K_{-i}$: 
\begin{eqnarray}
&&K_i(V_m) = \alpha_{0,i} e^{q_{\rm forward} V_{\rm m}/26.7},\\
&&K_{-i}(V_m) = \beta_{0,i} e^{-q_{\rm backward} V_{\rm m}/26.7},
\end{eqnarray}
where $\alpha_{0,0}$ through $\alpha_{0,4}$ are 5.5, 8,0, 2.0, 884, 900 s$^{-1}$, and $\beta_{0,0}$ through $\beta_{0,4}$ are 8.669, 1.127, 0.0252, 1.013, 0.1257 ms$^{-1}$, respectively. The rate constants $q_{\rm forward}$ and  $q_{\rm backward}$ are 1.16 and 1.94. 

\subsection{Regulation of the chemical potentials}

\indent\indent As described in the previous section, the continuous-time Markov process for the voltage-sensitive calcium channel is applied with the membrane voltage and the intracellular calcium concentration adjacent to each ion channel. The chemical potential for Ca$^{2+}$ is modulated in the on/off way with the channel state variable $S$ from the Markov process. In the two-dimensional scheme, the chemical potential of calcium is \emph{locally} regulated. Let the width of the ion channel be $w_{\rm ch}$. For the $j^{\rm th}$ ion channel in the state $S_j$, with the Lagrangian parameter $s_j$ for the location of the center of the channel, we make spatial regulation of the chemical potential in the following way:
\begin{equation}
A_{{\rm Ca}^{2+}}(s,t) = \left\{ \begin{array}{ll}
A_{{\rm Ca}^{2+}, {\rm open}}	& S_j = {\rm O},  s \in [s_j - w_{\rm ch}/2, s_j + w_{\rm ch}/2]\\
A_{{\rm Ca}^{2+}, {\rm closed}} & {\rm otherwise} \end{array} \right.
\end{equation}
where $A_{{\rm Ca}^{2+}, {\rm closed}}$ and  $A_{{\rm Ca}^{2+}, {\rm open}}$ are specified chemical potential amplitudes to make the membrane mostly impermeable or semi-permeable to calcium ions (Table 2). For the domain of the membrane without ion channels, the chemical potential for calcium is fixed. The regulation for calcium-activated potassium channels is mostly the same, but the regulated chemical potential is specified to potassium ions.

\subsection{Numerical implementation}

 The Poisson equation, Eq. (\ref{eq:poisson1}) is solved by Fourier transformation with electrical density given by the computed concentration for each ionic species. The electrodiffusion equations (Eqs. \ref{eq:continuity} and \ref{eq:flux}) are solved for the concentration of each ionic species by the backward Euler method with a second-order Godunov upwind method (Lee et al. 2010). We employ the Krylov subspace iterations (GMRES) using the PETSc library (Balay et al. 1997, Balay et al. 2019a,b) preconditioned by the algebraic multigrid methods in the \emph{hypre} library
(HYPRE, Falgout and Yang 2002). The numerical algorithm for the ion channel gating is based on the Monte Carlo method, determining whether to transit, and secondly the state to transit if needed. For the details, see Lee et al. 2013.\\

\section{RESULTS AND DISCUSSION} \label{results_discussion}

\indent\indent For the two-dimensional simulation, a periodic square domain with dimensions 2 $\mu$m $\times$ 2 $\mu$m is covered by a Cartesian grid containing 256 $\times$ 256 points with the uniform grid size $h$ in each direction. The model dendritic spine has a diameter 1 $\mu$m and is centered within the computational domain. A timestep $\Delta t$ = 30 $\mu$s is used for all computations, including both the electrodiffusion and the Markov process that controls the opening and closing of membrane channels. The electrical potential $\varphi$, the chemical potential $\psi_i$ and the concentration $c_i$ for $i^{\rm th}$ ionic species are defined on the cell-centered grids. For the support of the chemical potential kernel, $w = 6h$ is prescribed. 

The membrane changes its permeability to Ca$^{2+}$ and K$^+$ in a voltage- and calcium-dependent manner according to the continuous-time Markov process described above with spatially localized ion channels. For the simulation, five voltage-sensitive calcium channels and five calcium-activated potassium channels are uniformly distributed on the head's upper half-circle, representing a postsynaptic density (PSD) with 15 nm distance between two types of channels as a complex. These complexes are labeled indices from 1 to 5 counter-clockwise. Four ion species and background charges are considered with initial concentrations different between extracellular and intracellular domains, as shown in Table (1).

The membrane is stochastically permeable to calcium by the voltage-sensitive calcium channel. The amount of influx through the channel is regulated by controlling the chemical potential with two parameters, $A_{{\rm Ca}^{2+}, {\rm open}}$ and $A_{{\rm Ca}^{2+}, {\rm closed}}$ as prescribed in Table (2). Similarly, the membrane is stochastically permeable to potassium by the calcium-activated potassium channel. The amount of influx through the channel is regulated by controlling the chemical potential with two parameters, $A_{{\rm K}^{+}, {\rm open}}$ and $A_{{\rm K}^{+}, {\rm closed}}$.

\subsection{Stochastic calcium influxes by a train of back-propagating action potential waves: non-local feedback loop between two-types of channels }

\begin{figure}[h]
\centering
\includegraphics[angle=0, width=1.0\textwidth]{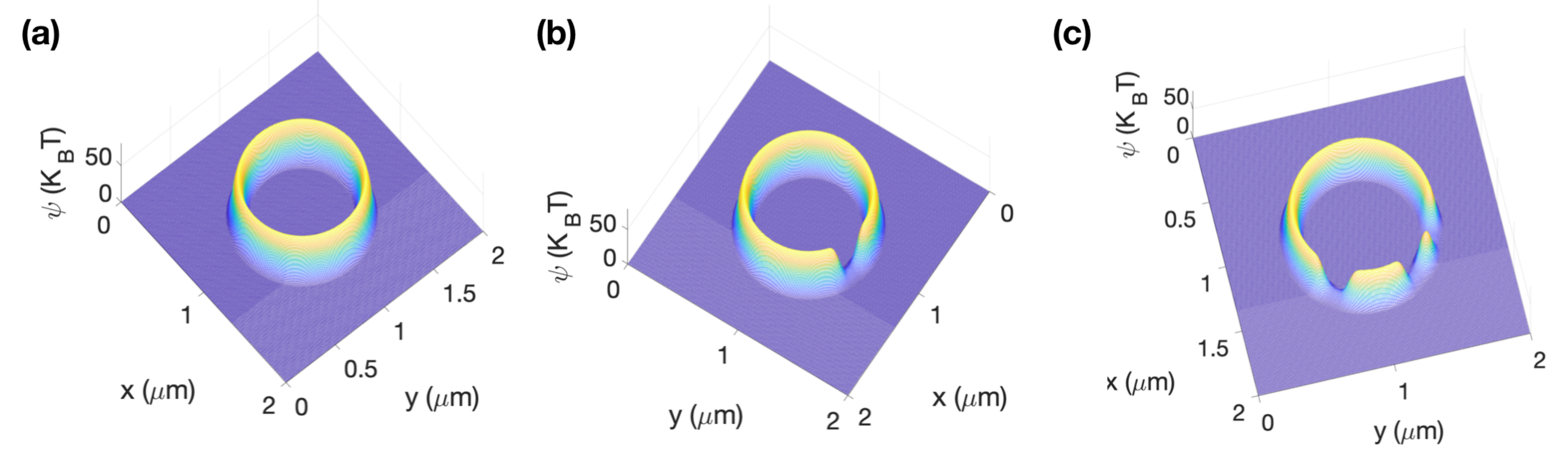}
\caption{\textbf{Chemical potential distribution in position and in the course of ion channel gating on 2D}: it is placed along the boundary. The support of the chemical potential kernel is a square that measures 24 mesh widths  $\times$ 24 mesh widths. The chemical potential is locally and stochastically regulated at the domain of ion channels. (a) The chemical potential for calcium is high, i.e., five ion channels are all closed. (b) and (c) The chemical potentials for calcium are low around 1 and 3 ion channels.}
\label{fig:cpotential}
\end{figure}

\begin{figure}[h]
\begin{center}
\includegraphics[angle=0, width=0.8\textwidth]{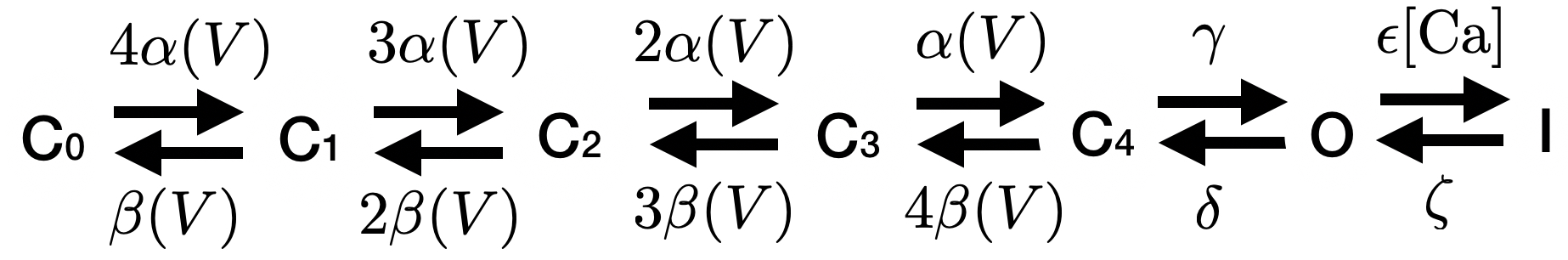}
\caption{\textbf{Markov chain of the voltage-sensitive calcium channel}: The calcium channel has four subunits. In the state of $C_i$, $i-1$ subunits are open. In the state of $\rm I$, the channel is in inactivation and closed. The ion channel is open only when it stays on the state of $\rm O$. The rate constants for the opening and closing of each subunit are denoted by $\alpha$ and $\beta$. Courtesy of Cox \cite{cox}}. \label{markov_CaV}
\end{center}
\end{figure}

\begin{figure}[h]
\begin{center}
\includegraphics[angle=0, width=1.0\textwidth]{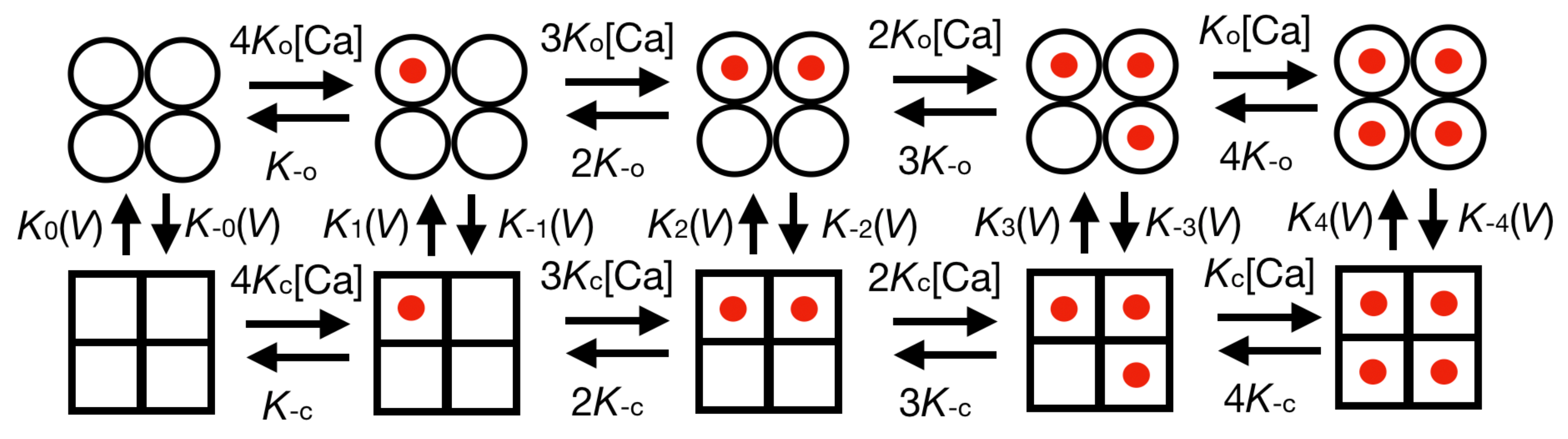}
\caption{\textbf{Markov chain of the calcium-activated potassium channel}: The channel has four subunits, and the calcium-binding rate to each subunit is calcium-dependent. The rate constants for the opening and closing in the five different calcium bound states are denoted by $k_i$ and $k_{-i}$, each dependent on the membrane voltage. Courtesy of Cox \cite{cox}}. \label{markov_BK}
\end{center}
\end{figure}

The depolarization of the membrane is implemented by lowering the chemical potential of sodium to 42.74 K$_{\rm B}T$ at $t$=3 ms. The membrane is repolarized by raising the sodium chemical potential to the initial level, 53.43 K$_{\rm B}T$ at $t$=21 ms. With the membrane voltage in the depolarization range, there comes calcium ionic influx from the extracellular domain.  The locally regulated chemical potential of Ca$^{2+}$ around the area of ion channels is shown in Fig. \ref{fig:cpotential} with its dynamic changes in the amplitude.

\begin{figure}[h] 
\begin{center}
$\begin{array}{c@{\hspace{0.1in}}c}
\includegraphics[angle=0, width=1.0\textwidth]{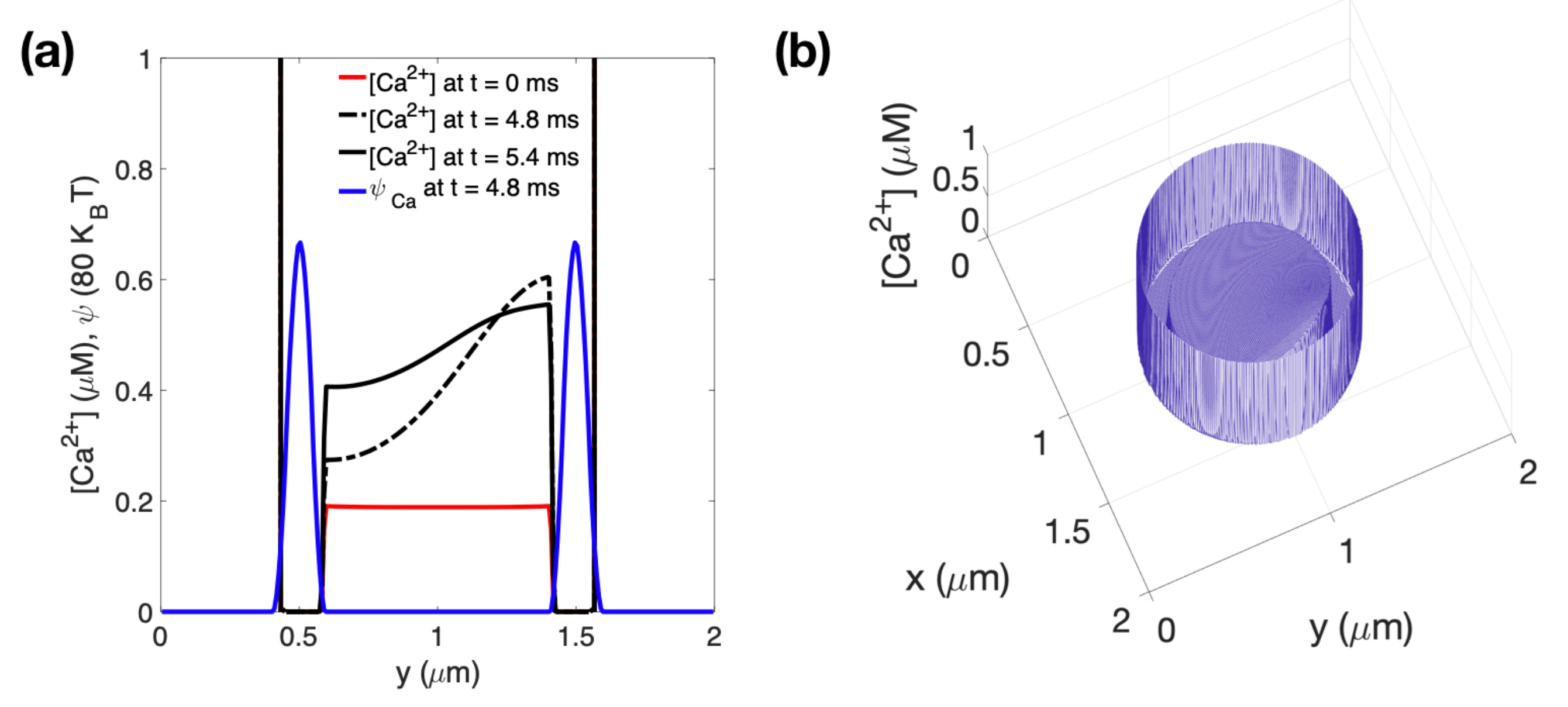}
\end{array}$
\caption{\textbf{Concentration distribution with ion channels open on 2D}; (a) The calcium concentration distribution in y-section at $t$=0 ms (red), $t$=4.8 ms (black-dotted), $t$=5.4 ms (black-solid). The chemical potential distribution is also shown (blue) at $t$=4.8 ms. (b) The calcium concentration is zoomed in to resolve the low-level concentration in $\mu$M in the intracellular domain at $t$=4.8 ms.} \label{fig:2D:ca}
\end{center}
\end{figure}

\begin{figure}[h] 
\begin{center}
$\begin{array}{c@{\hspace{0.1in}}c}
\includegraphics[angle=0, width=1.0\textwidth]{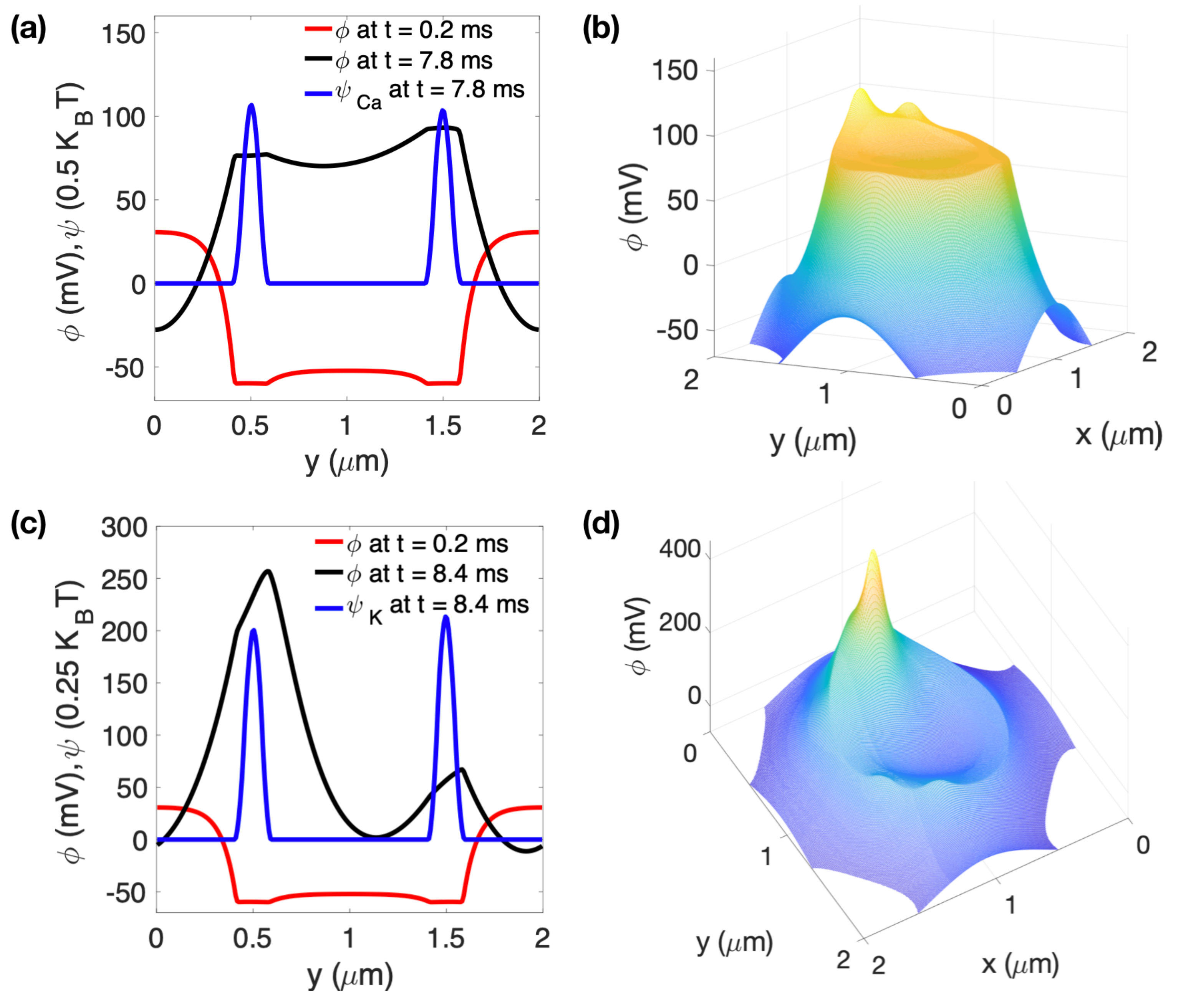}
\end{array}$
\caption{\textbf{Electrical potential distribution on 2D}; (a) The electrical potential in the y-section at $t$=0.2 ms (red) and $t$=7.8 ms (black). The chemical potential for the calcium ions is also shown in y-section at $t$=7.8 ms (blue). (b) The electrical potential on the 2D domain at $t$=7.8 ms. (c) The electrical potential in the y-section at $t$=0.2 ms (red) and $t$=8.4 ms (black). The chemical potential for potassium is also shown in y-section at $t$=8.4 ms (blue). (d) The electrical potential on the 2D domain at $t$=8.4 ms.} \label{fig:2D:epo}
\end{center}
\end{figure}

The calcium concentration distributions in the y-section are shown when calcium ions are flowed in, and influenced by the electrodiffusion at $t$=4.8 ms and at $t$=5.4 ms (Fig. \ref{fig:2D:ca}a). The calcium concentration distribution on the 2D domain is also shown zoomed in at $t$=4.8 ms in Fig. \ref{fig:2D:ca}(b). The electrical potential distributions in the y-section and the 2D domain are shown at $t$=7.8 ms and $t$=8.4 ms (Fig. \ref{fig:2D:epo}). At $t$=7.8 ms, two voltage-sensitive calcium channels are open, and the associated chemical potential is locally lowered. Accordingly, the membrane is more depolarized locally, showing two spikes. At $t$=8.4 ms, a calcium-activated potassium channel is open, and the corresponding chemical potential is locally decreased. There follows a considerable upstroke of electrical potential adjacent to the open BK channel, and the expected membrane repolarization does appear far away \emph{non-locally}, that is an unexpected result.  

\begin{figure}[h] 
\begin{center}
$\begin{array}{c@{\hspace{0.1in}}c}
\includegraphics[angle=0, width=1.0\textwidth]{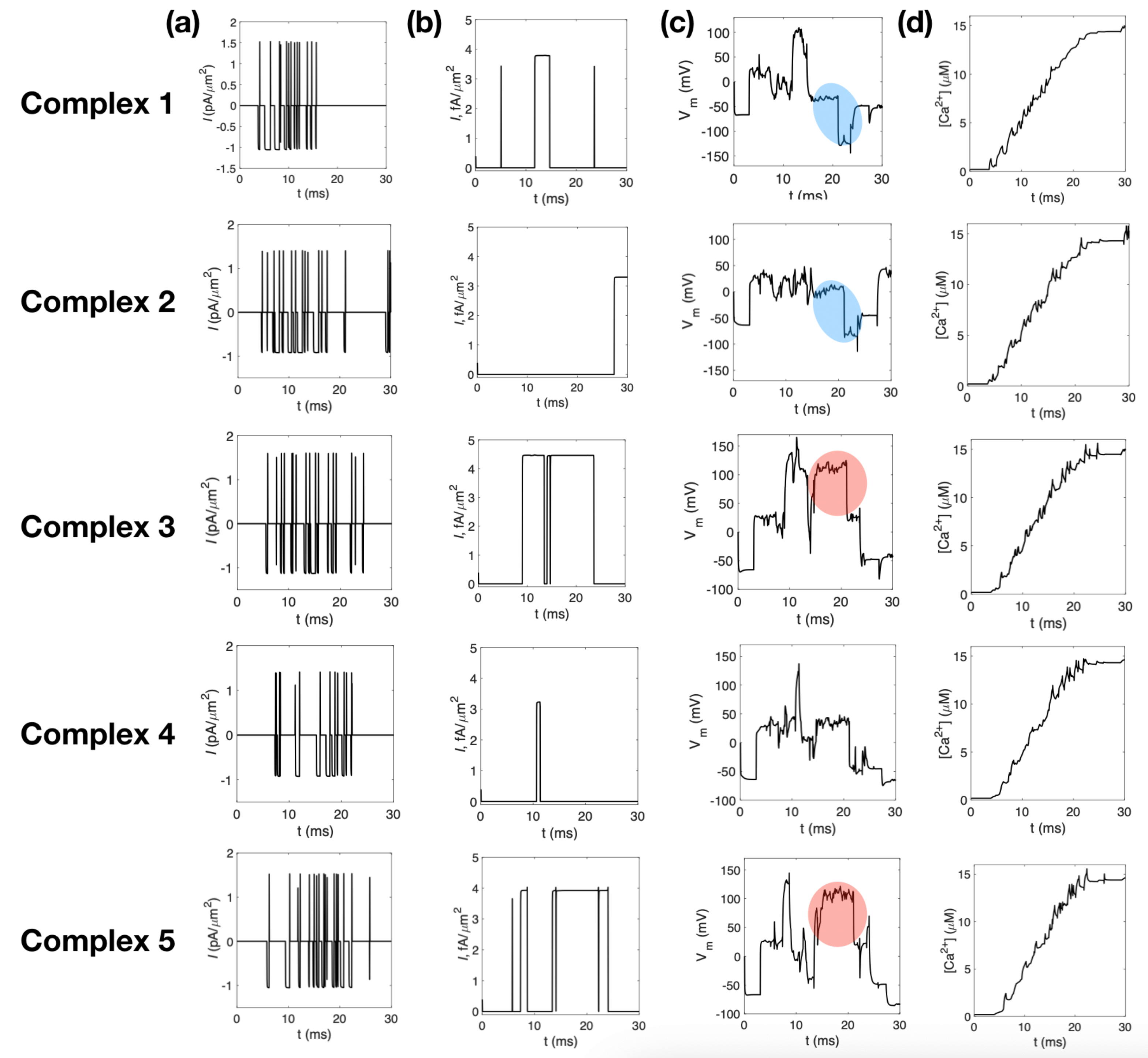}
\end{array}$
\caption{\textbf{Voltage-sensitive calcium channel and calcium-activated potassium channel activities}; The columns represent (a) calcium ionic currents by the voltage-sensitive calcium channels, (b) potassium ionic currents by the calcium-activated potassium channels, (c) membrane voltage, and (d) intracellular calcium concentration adjacent to each voltage-sensitive calcium channel. The rows represent five individual complexes of the voltage-sensitive calcium channel and the calcium-activated potassium channel. These electrophysiological activities are from the non-uniform distribution of the channel complexes.} \label{gating}
\end{center}
\end{figure}

The ionic current and membrane voltage in the time course of the channel gating are presented in Fig. \ref{gating}. When the membrane is depolarized, and the channels are open, we observe the calcium currents. Interestingly even when the membrane is repolarized, it takes a while for the calcium channels are all closed. The stochastic calcium-activated potassium channel 1 is shortly open around $t$=4 ms before the intracellular calcium concentration is elevated enough. The evident response of the calcium-activated potassium channel after the calcium inflow by the voltage-sensitive calcium channels is shown for the complexes 3 and 5 with some persistent outflow of the potassium ionic current.  Here one interesting aspect of membrane voltage change is that local membrane voltages are {\it increased} around the complexes 3 and 5 (red circled), and {\it decreased} around the complexes 1 and 2 (blue circled). This shows that the ionic current from the BK ion channel gating is influential \emph{non-locally} with the fast redistribution of the electrical potential by the electrodiffusion, as shown in Fig. \ref{fig:2D:epo}(c) and \ref{fig:2D:epo}(d). 

\subsection{Differential calcium handling by channel distributions}

We have tested whether different distributions of the multiple channels generate different calcium influxes, and whether the inhibition of the BK channel drives more calcium influx. We compare the averaged intracellular calcium concentration at $t$=30 ms for the non-uniformly distributed case with channel complexes uniformly distributed in the whole dendritic spine head (Fig. 7, first two datasets). The two-sample $t$-test shows a significantly higher intracellular calcium level for the non-uniformly distributed case. It is supposed that calcium influx induces membrane depolarization, and this provides positive feedback to adjacent voltage-sensitive calcium channels for them to have a higher probability of opening channels. This interaction is thought to be stronger than repolarizing effects from adjacent calcium-activated potassium channels. When the channel is open, the intracellular calcium concentration around a voltage-sensitive calcium channel is elevated, and the adjacent calcium-activated potassium channel gradually responds by opening it (Figs. 5c and 5d). What happens here is local depolarization and non-local repolarization. After that, this might enhance the coupled voltage-sensitive calcium channel to stay in an open state. The only action of channel inhibition is by calcium-sensitive inactivation in the channel itself.

We also compare the intracellular calcium concentrations with two cases of the BK channels expressed with their non-uniform distribution or knocked out (Fig. 7, second and third datasets). The paired-sample $t$-test shows a significant elevation of calcium influx by the BK channel inhibition or the knock-out. The significant difference in the intracellular calcium concentration shows the inhibitory mechanism on calcium influx through interaction between voltage-sensitive calcium channels and calcium-activated potassium channels.
\section{Conclusions}

\indent\indent By applying \textit{the immersed chemical potentials} on the two-dimensional domain, we realize the regulation of membrane permeability to each ionic species and, consequently, the ion selectivity of the membrane. With the continuous-time Markov process in the stochastic feature of the voltage-sensitive ion channel, the ion channel gating and the corresponding ionic current are observed depending on the membrane voltage. With discretely placed ion channels,  individual ion channel gating is observed with spatiotemporally changing chemical potential distributions. Compared to the point cell modeling, this kind of approach makes it available for us to consider the spatially \emph{non-uniform} distribution of ion channels and their spatial effects on the physiology. Accordingly, we can implement the change in the ion channel density in the sense of synaptic plasticity.

The \emph{non-local} interaction between voltage-sensitive calcium channels and calcium-activated potassium channels is thought to be captured by treating the electrophysiology with electrodiffusion in nano- to micro-scales, and undoubtedly not feasible by drift-diffusion. At the same time, positive feedback among \emph{local} channel complexes for calcium influx is also predicted for further validation by experiments. The back-propagating action potential was treated by the one-time event of phasic trains. In real neuronal activities, the frequency of this kind of back-propagating action potential spikes may induce differential responses from calcium-activated potassium channels (Kuznetsov et al. 2006, Hage and Khaliq 2015). We expect to reconstruct the three-dimensional electrodiffusion (Mori 2007, Cartailler and Holcman 2019), which is straightforward due to the Cartesian grids where the chemical potentials are immersed and the electrodiffusion is computed.

\section*{Acknowledgments}
The author appreciates careful comments from Charles Peskin.


%
%


\begin{thebibliography}{}

\bibitem{Balay} S. Balay, V. Eijkhout, W.D. Gropp, L.C. McInnes, B.F. Smith, Efficient management of parallelism in object-oriented numerical software libraries, in: E. Arge, A.M. Bruaset, H.P. Langtangen (Eds.), Modern Software Tools in Scientific Computing, Birkhauser Press, pp. 163-202 (1997)

\bibitem{Balay2} S. Balay, S. Abhyankar, M.-F. Adams, J. Brown, P. Brune, K. Buschelman, L. Dalcin, A. Dener, V. Eijkhout, W.D. Gropp, D. Karpeyev, D. Kaushik, M.G. Knepley, D.-A. May, L.C. McInnes, B.F. Smith, S. Zampini, H. Zhang, PETSc Users Manual, Technical Report, ANL-95/11 Revision 3.12, Argonne National Laboratory (2019)

\bibitem{Balay3} S. Balay, S. Abhyankar, M.-F. Adams, J. Brown, P. Bruen, K. Buschelman, L. Dalcin, A. Dener, V. Eijkhout, W.D. Gropp, D. Karpeyev, D. Kaushik, M.G. Knepley, L.C. McInnes, R.T. Mills, T. Munson, K. Rupp, P. Sanan, B.F. Smith, S. Zampini, H. Zhang, PETSc, http://www.mcs.anl.gov/petsc (2019)

\bibitem{berkefeld} H. Berkefeld, C.A. Sailer, W. Bildl, V. Rohde, J.-O. Thumfart, S. Eble, N. Klugbauer, E. Reisinger, J. Bischofberger, D. Oliver, H.-G. Knaus, U. Schulte, B. Fakler, BK${_{\rm Ca}}$-CaV channel complexes mediate rapid and localized Ca$^{2+}$-activated K$^+$ signaling, {\em Science}, 314, 615-620 (2006)

\bibitem{bloodgood} B.L. Bloodgood and B.L. Sabatini, Nonlinear Regulation of Unitary Synaptic Signals by CaV2.3 Voltage-Sensitive Calcium Channels Located in Dendritic Spines, {\em Neuron}, 53, 249-260 (2007)

\bibitem{bloodgood2} B.L. Bloodgood and B.L. Sabatini, Ca$^{2+}$ signalling in dendritic spines, {\em Curr. Opin. Neurobiol.}, 17, 345-351 (2007)

\bibitem{bonhoeffer} T. Bonhoeffer and R. Yuste, Spine motility: phenomenology, mechanisms, and function, {\em Neuron}, 35, 1019-1027 (2002)

\bibitem{bosch} M. Bosch and Y. Hayashi, Structural plasticity of dendritic spines, {\em Curr. Opin. Neurobiol.}, 22, 1-6 (2011)

\bibitem{cartailler} J. Cartailler and D. Holcman, Steady-state voltage distribution in three-dimensional cups-shaped funnels modeled by PNP, {\em J. Math, Biol.}, 79, 155-185 (2019)

\bibitem{cox} D.H. Cox, Modeling a Ca$^{2+}$ channel/BK$_{\rm Ca}$ channel complex at the single-complex level, {\em Biophy, J.}, 107, 2797-2814 (2014)

  \bibitem{FalgoutYang02}
R.~D. Falgout, U.~M. Yang, {\sl hypre}: a library of high performance
  preconditioners, in: P.~M.~A. Sloot, C.~J.~K. Tan, J.~J. Dongarra, A.~G.
  Hoekstra (Eds.), Computational Science - ICCS 2002 Part III, Vol. 2331 of
  Lecture Notes in Computer Science, Springer-Verlag, 2002, pp. 632--641, also
  available as LLNL Technical Report UCRL-JC-146175.
  
\bibitem{griffith} T. Griffith and K. Tsaneva-Atanasova and J.R. Mellor, Control of Ca$^{2+}$ influx and calmodulin activation by SK-channels in dendritic spines, {\em PLoS Comput. Biol.}, 12, e1004949 (2016)

\bibitem{gutzmann} J.J. Gutzmann and L. Lin and D.A. Hoffman, Functional coupling of Cav2.3 and BK potassium channels regulates action potential repolarization and short-term plasticity in the mouse hippocampus, {\em Front. Cell. Neurosci.}, 34, 5261-5272 (2019)

\bibitem{hage} T.A. Hage and Z.M. Khaliq, Tonic firing rate controls dendritic Ca$^{2+}$ signaling ad synaptic gain in substantia nigra dopamine neurons, {\em J. Neurosci.}, 35, 5823-5836 (2015)

\bibitem{hansel} C. Piochon, M. Kano, and C. Hansel, LTD-like molecular pathways in developmental synaptic pruning, {\em Nat. Neurosci.}, 19, 1299-1310 (2016)

\bibitem{harris} K.M. Harris and S.B. Kater, Dendritic spines: cellular specializations imparting both stability and flexibility to synaptic function, {\em Annu. Rev. Neurosci.}, 17, 341-371 (1993)

\bibitem{he} S. He, Y.-X. Wang, R.S. Petralia, S.D. Brenowitz, Cholinergic modulation of large-conductance calcium-activated potassium channels regulates synaptic strength and spine calcium in cartwheel cells of the dorsal cochlear nucleus, {\em J. Neurosci.}, 34, 5261-5272, (2014)

\bibitem{holcman} D. Holcman and R. Yuste, The new nanophysiology: regulation of ionic flow in neuronal subcompartments, {\em Nat. Rev. Neurosci.}, 16, 685-692 (2015)

\bibitem{hypre-web-page} HYPRE: High performance preconditioners, http://www.llnl.gov/CASC/hypre.
  
\bibitem{kim} I.H. Kim, M.A. Rossi, D.K. Aryal, B. Racz, N. Kim, A. Uezu, F. Wang, W.C. Wetsel, R. Weinberg, H. Yin, and S.H. Soderling, Spine pruning drives antipsychotic-sensitive locomotion via circuit control of striatal dopamine, {\em Nat. Neurosci.}, 18, 883-891 (2015)

\bibitem{koleske} A.J. Koleske, Molecular mechanisms of dendrite stability, {\em Nat. Rev. Neurosci.}, 14, 536-550 (2013)

\bibitem{kuznetsov} A.S. Kuznetsov, N.J. Kopell, and C.J. Wilson, Transient high-frequency firing in a coupled-oscillator model of the mesencephalic dopaminergic neuron, {\em J. Neurophysiol.}, 95, 932-947 (2006)

\bibitem{lamprecht} R. Lamprecht, C.R. Farb, S.M. Rodrigues, and J.E. LeDoux, Fear conditioning drives prolifin into amygdala dendritic spines, {\em Nat. Neurosci.}, 9, 481-483 (2006)

\bibitem{lee07} P. Lee, The immersed boundary method with advection-electrodiffusion, Ph.D. thesis, Courant Institute of Mathematical Sciences, New York University (2007)

\bibitem{lee_griffith_peskin10} P. Lee and B.E. Griffith and C.S. Peskin, The immersed boundary method for advection-electrodiffusion with implicit timestepping and local mesh refinement, {\em J. Comp. Phys.}, 229, 5208-5227 (2010)

\bibitem{lee_sobie_peskin13} P. Lee, E.A. Sobie, C.S. Peskin, Computer simulation of voltage sensitive calcium ion channels in a dendritic spine, {\em J. Theo. Biol.}, 338, 87-93 (2013)

\bibitem{lujan} R. Lujan, C. Aguado, F. Ciruela, X.M. Arus, A. Martin-Belmonte, R. Alfaro-Ruiz, J. Martinez-Gomez, L. de la Ossa, M. Watanabe, J.P. Adelman, R. Shigemoto, Y. Fukazawa, SK2 channels associate with mGlu$_{1\alpha}$ receptors and Ca$_{V}$2.1 channnels in purkinje cells, {\em Front. Cell. Neurosci.}, 12, 1-16 (2018)

\bibitem{matus} I. Brunig and S. Kaech and H. Brinkhaus and T. G. Oertner and A. Matus, Influx of extracellular calcium regulates actin-dependent morphological plasticity in dendritic spines, {\em Neuropharmacology}, 47, 669-676 (2004)

\bibitem{mori} Y. Mori and J. W. Jerome and C.S. Peskin, A three-dimensional model of cellular electrical activity, 
{\em Bull. Inst. Math. Acad.}, 2, 367-390 (2007)

\bibitem{peskin} C.S. Peskin, The Immersed Boundary Method, {\em Act. Num.}, 11, 479-517 (2002)

\bibitem{peskin2} C.S. Peskin, Mathematical Aspects of Neurophysiology, Lecture Note, Courant Institute of Mathematical Sciences, New York University (2000)

\bibitem{rice} R.A. Rice, E.E. Spangenberg, H. Yamate-Morgan, R.J. Lee, R.P.S. Arora, M.X. Hernandez, A.J. Tenner, B.L. West, K.N. Green, Elimination of microglia improves functional outcomes following extensive neuronal loss in the hippocampus, {\em J. Neurosci.}, 35, 9977-9989 (2015)

\bibitem{sah} P. Sah and E.S.L. Faber, Channels underlying neuronal calcium-activated potassium currents, {\em Progress in Neurobiology}, 66, 345-353 (2002)

\bibitem{sheng} M. Sheng and M. Kim, Postsynaptic Signaling and Plasticity Mechanisms, {\em Science}, 298, 776-780 (2002)

\bibitem{shepherd} G.M. Shepherd, The Dendritic Spine: A Multifunctional Integrative Unit, {\em J. Neurophysiol.}, 75, 2197-2210 (1996)

\bibitem{svoboda2} B.L. Sabatini and K. Svoboda, Analysis of calcium channels in single spines using optical fluctuation analysis, {\em Nature}, 408, 589-593 (2000)

\bibitem{svoboda3} R. Yasuda and B.L. Sabatini and K. Svoboda, Plasticity of calcium channels in dendritic spines, {\em Nat. Neurosci.}, 6, 948-955 (2003)

\bibitem{tada} T. Tada and M. Seung, Molecular mechanisms of dendritic spine morphogenesis, {\em Curr. Opin. Neurobiol.},16, 95-101 (2006)






%
%

\end{thebibliography}

\clearpage

\clearpage
\clearpage
\clearpage
\clearpage
\clearpage
\begin{figure}[h]
\begin{center}
$\begin{array}{c@{\hspace{0.1in}}c}
\includegraphics[angle=0, width=0.5\textwidth]{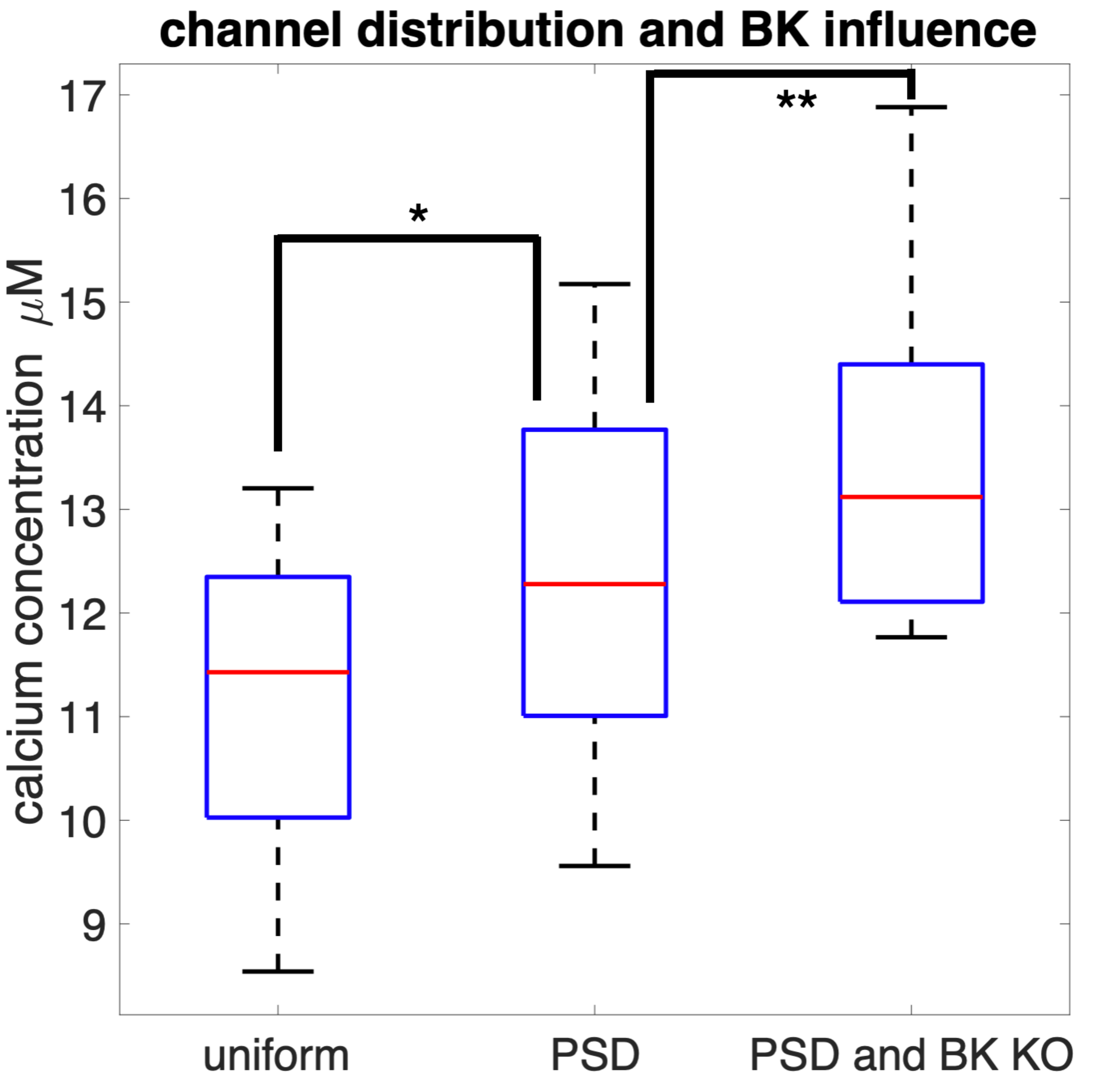}
\end{array}$
\caption{\textbf{Calcium ion concentration comparison}; ten samples are collected for the comparison of the averaged intracellular calcium ion concentrations for three cases, 1) the ion channel complexes are uniformly distributed, 2) the ion channel complexes are non-unformed distributed, 3) the ion channels complexes are non-uniformly distributed, and the BK channels are knocked out. The activity of the BK channel functions as inhibiting the calcium inflow. Non-uniform distribution of the ion channel complexes induces higher intracellular calcium concentrations. * p $<$ 0.05 two-sample $t$-test, ** p $<$ 0.05 paired-sample $t$-test \label{calcium:BK:comparison}} 
\end{center}
\end{figure}
\clearpage
\begin{table}[h]
\caption{\textbf{Initial concentrations in all simulations (mM)}. X$^-$ denotes the fixed background charge and the concentrations stated for X$^-$ refer to the concentration of the background charges, not the concentration of the molecules that carry the background charges. \label{table:concentration}}
\begin{center}
\begin{tabular}{|c|c|c|}
\hline
Ion species & extracellular concentration & intracellular concentration\\
\hline
Ca$^{2+}$ & 2.0 & 0.0002\\
Cl$^-$ & 150 & 13\\
Na$^+$ & 150 & 15\\
K$^+$ & 5 & 100\\
X$^-$ & 9 & 102.0004\\
\hline
\end{tabular}
\end{center}
\end{table}
\begin{table}[h]
\caption{Chemical potential (K$_{\rm B}$T) prescribed for each ionic species.  \label{table:cpo}}
\begin{center}
\begin{tabular}{|c|c|c|}
\hline
Ion species & chemical potential high & chemical potential low\\
\hline
Ca$^{2+}$ & 53.43 & 8.014 \\
Cl$^-$ & 45.41 & 45.41\\
Na$^+$ & 53.43 & 42.74\\
K$^+$ & 53.43 & 18.70\\
X$^-$ & 53.43 & 53.43\\
\hline
\end{tabular}
\end{center}
\end{table}
%


\end{document}